\documentclass[fleqn,usenatbib]{mnras}

\usepackage{newtxtext,newtxmath}

\usepackage[T1]{fontenc}

\DeclareRobustCommand{\VAN}[3]{#2}
\let\VANthebibliography\thebibliography
\def\thebibliography{\DeclareRobustCommand{\VAN}[3]{##3}\VANthebibliography}

\usepackage{graphicx}	
\usepackage{amsmath}	

\usepackage{placeins}        
\usepackage{threeparttable}  

\usepackage{xcolor}          
\usepackage[normalem]{ulem}

\newcommand{\tbar}{\,T\textsubscript{1 bar}}

\title[]{The atmospheric vertical structure of Uranus and Neptune from thermochemical models: the impact of model assumptions}

\author[T. Douçot et al.]{
Thomas Douçot,$^{1}$\thanks{E-mail: thomas.doucot@uzh.ch}
Ravit Helled,$^{1}$
Daniel Kitzmann,$^{2,3}$
Ricardo Hueso,$^{4}$
and Audrey Vorburger$^{2}$
\\
$^{1}$Institut für Astrophysik, Universität Zürich, 8057 Zurich, Switzerland\\
$^{2}$Space Research and Planetary Sciences, Physics Institute, University of Bern, 3012 Bern, Switzerland\\
$^{3}$Center for Space and Habitability, University of Bern, 3012 Bern, Switzerland\\
$^{4}$Planetary Sciences Group, Dpto. Física Aplicada,
Escuela de Ingeniería de Bilbao, UPV/EHU, Plaza Ingeniero Torres Quevedo, 1 
48013 Bilbao, Spain
}

\date{Accepted XXX. Received YYY; in original form ZZZ}

\pubyear{2026}

\begin{document}
\label{firstpage}
\pagerange{\pageref{firstpage}--\pageref{lastpage}}
\maketitle

\begin{abstract}
The composition and temperature-pressure profile of the atmospheres of Uranus and Neptune are not well-determined. As observational data are limited, we often rely on chemical equilibrium computations to infer atmospheric abundances and cloud formation. The inferred atmospheric structures, however, strongly depend on several fundamental assumptions such as the elemental abundances and ratios, the condensation properties of the assumed species, or a reference temperature for the adiabatic structure.  In this study we investigate the effects of different metallicities (1 to 80 solar), element ratios (C/O and S/N, from 0.1 to 2 and 0.19 to 1.6) and 1 bar temperatures (66 to 86 K) on the vertical structure of ice giant atmospheres. In particular,  we use the chemical equilibrium code \texttt{FastChem} to derive mixing ratios and cloud structures for CH$_4$, NH$_3$, H$_2$S, H$_2$O and NH$_4$SH. We find that the models are very sensitive to the assumed parameters, yielding drastically different possible atmospheric structures. For the cases considered here, we find that mixing ratios and cloud deck altitudes can vary by more than an order of magnitude. Additionally, thermal profiles can differ by several tens of kelvins due to composition and 1-bar temperature. We advise that future ground-based observations and a dedicated mission to Uranus and/or Neptune are required to better characterize the atmospheric structure and composition of ice giants.
\end{abstract}

\begin{keywords}
planets and satellites: atmospheres -- planets and satellites: composition -- planets and satellites: gaseous planets -- planets and satellites: Uranus -- planets and satellites: Neptune
\end{keywords}



\section{Introduction}

Uranus and Neptune, also referred to as the "ice giants", are the outermost planets in the Solar System. Unlike their gas counterparts Jupiter and Saturn, which have benefited from multiple space missions, these cold worlds have only been visited once by the Voyager 2 spacecraft. As a result, there are still many open questions regarding the formation, evolution, interiors, and atmospheric structures of Uranus and Neptune \citep{2020RSPTA.37890474H}. Since the discovery of the first exoplanet \citep{1995Natur.378..355M} which expanded the field of planetary sciences beyond the Solar System, it has appeared that planets with masses/sizes similar to those of Uranus and Neptune are ubiquitous across the Milky Way \citep{2020SSRv..216...18A}. These intermediate-mass planets constitute a prominent class of exoplanets, and their atmospheres can now be characterized  \citep{2024ApJ...961L..23B}. It is therefore clear that a deeper exploration of Uranus and Neptune is required. This is crucial for the understanding of the Solar System as well as for the characterization of the most common planetary type in the galaxy. Despite their importance, fundamental questions about Uranus and Neptune remain unsolved, including their atmospheric and bulk composition and their internal structure. As a result, the Origins, Worlds and Life NASA Decadal Survey identified a dedicated Uranus Orbiter and Probe (UOP) mission as the  highest priority for the 2023-2032 period \citep{NAP26522}.\\

Atmospheric characterization is essential to draw an accurate link with deeper processes and properties and therefore for constraining the planetary origin and thermal evolution \citep{2020RSPTA.37890474H, 2026enap....1...51H}.

Although there has been no dedicated space mission to
the ice giants yet, ground-based and space telescope  observations have provided valuable insights into these planets despite the inherent observational challenges.
Especially, an effort has been made to measure zonal winds \citep{2012AJ....143..150F, 2014Icar..237..211D, 2015Icar..258..192S} as well as to probe atmospheric composition and cloud structure \citep{2016Icar..271..418I, 2018NatAs...2..420I, 2019Icar..317..266S}. Recent observations from the James Webb Space Telescope provide temperatures and chemistry of the upper atmosphere as well as information about the upper hazes and cloud layers \citep{2025epsc.conf.1261R, 2025epsc.conf.1210R}. However, limitations in the spatial, spectral and temporal resolution of such observations prevent us from retrieving fundamental atmospheric properties such as the composition and vertical structure. Consequently, this complicates the characterization of both the atmosphere and interior, and their interplay. Therefore, currently, we must rely on simulations to infer the internal structure and evolution of Uranus and Neptune \citep{2024A&A...684A.191N, 2025A&A...704A.183M}.
Characterizing the atmospheres of Uranus and Neptune is challenging. These planets have cold, hydrogen-rich atmospheres, and are enriched with heavy-elements compared to the Sun \citep{2020SSRv..216...58C, 2020SSRv..216...18A, 2020RSPTA.37890474H, 2022ExA....54..975M, 2026enap....1...51H}. In these environments, many chemical species can condense, and affect the composition and thermal structure of the atmospheres. Clouds also present an observational challenge by obscuring the light for instruments observing in the visible and near-infrared wavelengths. As a result, our understanding of the planetary atmospheric structure from observations is typically restricted to the regions above the uppermost cloud layer, down to a few bars. Thus, determining the physical and chemical properties of the atmosphere below relies on theoretical models. 
The presence of condensible species also contributes to the stability of the atmosphere against convection, which has consequences on the dynamics, thermal profile, and energy transport. Cloud formation is a complex process that requires prior knowledge on the local dynamics and microphysical processes, such as nucleation and condensation. The complexity of the problem increases further when considering the many different species that can coexist with the gas phase. As a result, for simplicity, the investigation of the vertical composition and cloud structure in planetary atmospheres has traditionally been done assuming chemical equilibrium.  


Chemical equilibrium models provide a useful first-order framework for investigating the chemical and condensate structure of planetary atmospheres. They allow the inference of global one-dimensional, horizontally averaged atmospheric structures from a set of inputs, such as the elemental abundances and the thermal profile.  
Equilibrium calculations provide a thermodynamically consistent reference for identifying and quantifying disequilibrium processes. They also provide a direct framework for exploring the dependence of atmospheric composition and cloud formation on fundamental parameters such as elemental abundances and thermal structure. Because the bulk composition, heavy-element abundances, and thermal profiles of the atmospheres of Uranus and Neptune remain poorly constrained, it is essential to first characterize the range of atmospheric structures allowed by thermochemical equilibrium before considering additional sources of physical complexity.
In addition, chemical equilibrium models are often applied to Solar System giant planets, especially for predicting the occurrence of clouds \citep{1994Icar..110..117F, 2002Icar..155..393L, 2006ApJ...648.1181V}. For such planets, atmospheric models are usually derived through the properties of deep atmospheric gas parcels, raised adiabatically and assuming local chemical equilibrium \citep{1973Icar...20..465W, 1985rapm.book...17A}. These models require knowledge of the deep composition (that is, the present species and their abundances). There are several approaches to tackle the question of equilibrium composition, such as minimizing the Gibbs free energy of the system \citep{1958JChPh..28..751W} or solving the law of mass action and element conservation equations \citep{1946JChPh..14..563B, 1947JChPh..15..107B}. However, adding condensation to the equilibrium gas phase composition is complex and computationally expensive. In cold planetary atmospheres, a large variety of different species can condense. Although the amount of stable condensed species in equilibrium with the gas phase is restricted by Gibbs' phase rule, it remains  unknown which species actually condense. Deriving vertical atmospheric models with such methods can then be seen as a two component approach. First, comes the equilibrium composition of the gas phase. Then, the set of stable condensates is selected. Due to the large number of species that can condense, such a process can be performed iteratively, until the Gibbs' phase rule and element conservation with the gas are satisfied \citep{2024MNRAS.527.7263K}. 
In this work, we use chemical equilibrium framework to investigate the atmospheric structure of Uranus and Neptune over a broad range of plausible compositions and thermal profiles. We infer the distribution and condensation of main species: CH$_4$, NH$_3$, H$_2$S, H$_2$O and NH$_4$SH.
Previous studies assumed specific element abundances and ratios \citep[such as][]{1973Icar...20..465W, 2005SSRv..116..121A, 2020RSPTA.37890476H}, although these are rather uncertain in the case of Uranus and Neptune. Also, the thermal profiles for moist adiabats were inferred assuming saturation curves for individual species until each of them reached a fixed deep abundance. It is therefore important to  investigate the sensitivity of the atmospheric models to changes in the assumed composition and thermal structures. Rather than attempting to construct a fully self-consistent radiative–convective or disequilibrium model, we focus on the equilibrium solutions themselves and quantify how uncertainties in the elemental abundances and thermal structure affect the predicted vertical composition and cloud condensation levels. This approach allows us to identify the range of atmospheric structures that are compatible with current constraints and to establish a reference framework for future studies incorporating additional physical and chemical processes.
This paper is organized as follows. In Sec.~\ref{sec:Methods} we describe all the settings for our computations. The inferred atmospheric structures are presented in Sec.~\ref{sec:results}. In Sec.~\ref{sec:discussion}, we discuss the results. Our conclusions are summarized in Sec.~\ref{sec:conclusions}. 

\section{Building 1D atmospheric models}
\label{sec:Methods}
\subsection{Chemical equilibrium and cloud condensation}
To build vertical atmospheric models, we used the open-source chemical equilibrium code \texttt{FastChem}\footnote{\url{https://github.com/NewStrangeWorlds/FastChem}}, which is described in detail in \cite{2018MNRAS.479..865S, 2022MNRAS.517.4070S, 2024MNRAS.527.7263K, 2026arXiv260518264K}. It solves the mass action law for elements $j \in \mathcal{E}$ and species $i \in \mathcal{S} \setminus \mathcal{E}$, such as:
\begin{equation}
n_i = K_i(T) \prod_{j \in \mathcal{E}} n_j^{\nu_{ij}}, 
\end{equation}
where $n_i$ is the number density of species $i$, $K_i(T)$ its temperature-dependant equilibrium constant, $n_j$ the number density of atoms for element $j$ and $\nu_{ij}$ their stoichiometric coefficients. Elemental conservation, when considering both the gas phase and condensate species, is given by:
\begin{equation}
N_j = n_j + \sum_{i \in \mathcal{S} \setminus \mathcal{E}} \nu_{ji} n_i + \sum_{c \in \mathcal{C}_s} \nu_{cj} n_c
\end{equation}

where $N_j$ is the total number density of atomic nuclei for element $j$, and $n_c$ and $\nu_c$ are the number density and stoichiometric coefficients for species $c$ in a set of stable condensates $\mathcal{C}_s$, respectively. This set of equations allows for the recovery of the equilibrium composition of a gas phase, and the extraction of activity and stability criteria to compute the formation and rainout of a condensed phase. \texttt{FastChem} contains about 500 gas phase and condensate species for up to 80 elements. We focus on the main volatile species expected to condense in the atmospheres of Uranus and Neptune, namely CH$_4$, NH$_3$, H$_2$S, H$_2$O and NH$_4$SH. \texttt{FastChem} takes a thermal profile as input and scans it from the highest to the lowest pressure. At each pressure, temperature (P, T) point, it calculates the equilibrium composition for the gas phase, and iteratively selects the most stable set of condensates while satisfying the phase rule with the gas. It offers a rainout approximation, so the gas phase gets progressively depleted if a species begins to condense. Once a cloud layer forms, the condensate is removed from the overlying atmosphere and is therefore no longer available to participate in chemical reactions at the lower temperatures found higher in the atmosphere. This "rainout condensation chemistry" does not include processes such as precipitation settling flux. It differs from standard equilibrium chemistry, in which condensates remain in exchange with the gas phase and can continue to react at lower temperatures. We construct the atmospheric models by separately changing the assumed (i) atmospheric metallicity, (ii) C/O and S/N ratios and (iii) 1-bar temperature. The following subsections describe the models settings for each of these parameters, which are listed in Table~\ref{tab:mod_parameters}. 

\subsection{Atmospheric metallicity}
\label{subs:metallicity}

The first parameter we varied is the assumed atmospheric metallicity (hereafter, Z). In the Solar System, the heavy-element abundance in giant planet atmospheres seems to increase with orbital distance \citep{2022ExA....54..975M}. Therefore, we performed our calculations for a wide range of metallicities, ranging from 1 to 80 times solar (Z$_\odot$). Although an atmospheric metallicity of 1 Z$_\odot$ is not expected for Uranus and Neptune, it provides a useful reference case for illustrating the effects of increasing atmospheric enrichment. By default, \texttt{FastChem} offers several solar element abundance files. We ran our calculations using the present-day photospheric ones taken from \cite{2009ARA&A..47..481A}. We added a special treatment for NH$_3$. As discussed in \cite{1978Icar...34...10G,  2018NatAs...2..420I, 2019Icar..321..550I, 2019AJ....157..251T, 2021PSJ.....2....3M}, the lack of NH$_3$ in deep microwave and radio observations as well as the detection of H$_2$S in the upper atmosphere of Uranus and Neptune suggest a high S/N ratio (more than $\sim$5 times solar, i.e. S/N $\geq$ 1). NH$_3$ is also thought to react with H$_2$S to form a deep ammonium hydrosulphide (NH$_4$SH) cloud, such as H$_2$S(g) + NH$_3$(g) $\rightarrow$ NH$_4$SH(s). The atmosphere is then not thought to bear NH$_3$ clouds. As a result, for each metallicity, we depleted the nitrogen in the atmosphere assuming that it is a factor of ten lower than the other heavy-elements.


\subsection{Elemental ratios}
\label{subs:elemratio}

It is possible that the elemental ratios of different species in Uranus' and Neptune's atmospheres differ from the solar ones. Planet formation models suggest that the formation process would lead to other ratios \citep{2014ApJ...793....9A}. We therefore considered different C/O and S/N ratios. The C/O ratio is an important tracer of the evolution history of Uranus and Neptune, while the S/N ratio has a strong control over the expected deep H$_2$S + NH$_3$ chemistry, and therefore, the presence and detectability of these species in the upper atmosphere. In \texttt{FastChem}, elemental ratios are changed by increasing or decreasing the abundance of individual elements. We controlled the C/O ratio via the carbon abundance and the S/N ratio via the sulphur abundance. Our C/O and S/N ratios range from 0.1 to 2.0 and 0.19 to 1.6, respectively, for a fixed metallicity of 30 Z$_{\odot}$. Therefore, in the cases where $Z >$ 1 $Z_{\odot}$, our NH$_3$ depletion framework reduces the nitrogen abundance by a factor of 10. This depletion was not applied to the models with varied S/N ratios as this would shift the S/N ratio to values between 1.9 and 16. Consequently, these models would no longer probe the S/N $\approx 1$ regime, which is essential for accurately capturing the H$_2$S-NH$_3$ chemistry.

\subsection{1-bar temperature and thermal profile}
\label{subs:pt_profile}

Changing the 1-bar temperature (hereafter, \tbar) requires a re-calculation of the entire thermal profile. It is often assumed that the thermal lapse rate is driven by convective motion and follows an adiabat. Indeed, the thermal profile follows a dry adiabat when no condensible species is present in the atmosphere, or when the gas is unsaturated. This  prescription changes in the presence of condensible species (or vapor). When the atmosphere is saturated and a species condenses, the thermal profile follows a moist adiabat \citep{1995Sci...269.1697G, 2017A&A...598A..98L, 2018JAtS...75.1063L}. This formalism considers that in H$_2$-He rich atmospheres, condensation triggers two competing effects: on the one hand, latent heat is released upon condensation, which facilitates convection by increasing the buoyancy of the air parcel. On the other hand, it creates molecular weight gradients which stabilize the atmosphere against convection.
The thermal profiles we present were calculated following  \cite{2017A&A...598A..98L}, where the dry adiabat is given by:

\begin{equation}
\dfrac{d\ln{T}}{d\ln{P}} = \nabla^{}_{T} = \dfrac{R}{\mu c_p}, 
\end{equation}
where $R$ is the ideal gas constant, $\mu$ the mean molecular weight of the air parcel, and $c_p$ is the mean specific heat capacity of the air parcel. When the atmosphere is saturated, the moist adiabat is given by:

\begin{equation}
\dfrac{d\ln{T}}{d\ln{P}} = \nabla^{*}_{T} = \dfrac{R}{\mu c_p} \frac{\left(1 + \dfrac{q_s}{1-q_s}\dfrac{M_d L}{RT}\right)}{\left(1 + \dfrac{q_s}{1-q_s}\dfrac{L}{c_p T}\gamma_s\right)}, 
\end{equation} 
where $q_s (T)$ is the saturation mass mixing ratio of the vapor, and $L(T)$ its specific latent heat and $M_d$ the mean molar mass of the atmosphere's dry phase (here H$_2$-He). This method allows us to compute dry or moist adiabats in a gas mixture with a dry and a vapor phases, as function of the abundance of the vapor. In this study, we adopted H$_2$O as the reference species. This choice is motivated by its potentially large abundance in ice giant interiors \citep{2020RSPTA.37890477M, 2026arXiv260604510C}, its high latent heat of condensation, and its ability to generate significant molecular weight gradients in the atmosphere. Based on the abundance of H$_2$O, the thermal profile can be separated into the following different regimes:

\begin{enumerate}
    \item[$\bullet$]An upper, moist convective region: as long as the vapor condenses, the thermal gradient follows the moist adiabat $\nabla^{*}_{T}$. The atmosphere is assumed saturated, so the vapor mixing ratio is the saturation one.
    \item[$\bullet$] A deep, dry layer: when the vapor mixing ratio reaches a prescribed deep value $q_{i}$ and no longer condenses, the thermal profile follows the dry adiabat $\nabla_{T}$.
    \item[$\bullet$] If the vapor mixing ratio reaches a critical value $q_c$, convection is inhibited and heat is transported through a radiative gradient (see Sec.~\ref{subs:convinhibit} for further discussion), based on the planet's internal flux and atmospheric opacity such as:
    
    \begin{equation}
    \nabla_{r} = \dfrac{3}{16}\dfrac{P\kappa}{g}\dfrac{F_{int}}{\sigma T^4}
    \label{eq: rad-grad}, 
    \end{equation}
    where $\kappa$ is the Rosseland mean opacity for a gaseous H-He + H$_2$O mixture, $F_{int}$ is the planetary internal flux, $g$ is the gravitational acceleration and $\sigma$ is the Stefan-Boltzmann constant. For $\kappa$, we used the analytical fit described in \citep{2013ApJ...775...10V}, which includes a dependence to temperature and metallicity.
\end{enumerate}

This formalism as described in detail in \cite{2017A&A...598A..98L} provides a simplified description of moist convection in giant planet atmospheres, as it considers a single vapor. We  adopted H$_2$O as the dominant species controlling the moist adiabat,  another candidate is CH$_4$. However, the condensation of methane primarily affects the upper troposphere \citep{2020RSPTA.37890476H, 2022ExA....54.1027G} and therefore is expected to have a limited influence on profiles extending to several thousands of bars compared to H$_2$O. We note, however, that in reality, multiple species may coexist and condense. \cite{2018JAtS...75.1063L} demonstrated that interactions between simultaneously condensing vapors introduce additional cross terms in the moist adiabatic formulation, particularly for the H$_2$O–NH$_3$ system, highlighting the limitations of single-species approaches. However, NH$_3$ condensation has not yet been conclusively identified in the atmospheres of Uranus and Neptune, and the major condensible species are expected to condense at distinct pressure levels, except potentially for NH$_4$SH and H$_2$O. In addition, this formalism assumes an equilibrium condensation and therefore neglects cloud microphysical processes such as supersaturation, nucleation, condensate retention, and precipitation. These effects may modify the detailed cloud structure and local thermal gradients \citep{2022RemS...15..219P, 2024A&A...686A.131L}.
Nevertheless, this approach provides a reasonable first-order approximation for constructing thermal profiles suitable for equilibrium chemistry calculations, for a limited set of parameters. Our reference thermal profiles have a \tbar~of 76 K for Uranus and 72 K for Neptune, as determined by the Voyager 2 radio occultation measurements \citep{1987JGR....9214987L, 1990GeoRL..17.1733L}. We constructed profiles with \tbar~ranging from 66 to 86 K, for a fixed metallicity of Z = 30 Z$_\odot$. This corresponds to 76$\pm{10}$ K for Uranus and 72$\pm{10}$ K for Neptune. These temperature ranges intentionally exceed the uncertainties of the Voyager 2 measurements ($\pm$2 K). 
While these formal 1-bar temperature uncertainties are low, we note that the true uncertainty is expected to be higher as the formal error usually reflects only measurement and inversion uncertainties, while the retrieval depends on several atmospheric assumptions \citep[e.g.,][and references therein]{2022PSJ.....3..159G}. 
In addition, the range of  66--86 K we consider is meant to cover a range that is appropriate for both Uranus and Neptune. 
For pressures lower than 1 bar, we used the Voyager 2 profiles, and shifted them to match the corresponding \tbar. The sources for all the variables used to calculate the thermal profiles are listed in Appendix.~\ref{apdx:ptprofiles}.

\renewcommand{\arraystretch}{1.2}

\begin{table}
    \centering
    \caption{\label{tab:mod_parameters} \texttt{FastChem} settings summary.}
        \begin{threeparttable}
            \begin{tabular}{lcc}
                \hline\hline
                Parameter & Range & Reference value\\
                \hline
                Metallicity (Z$_\odot$) & [1 - 80] & 1 \tnote{(a)}\\
                C/O & [0.1 - 2.0] & 0.55 \tnote{(a)}\\
                S/N & [0.19 - 1.6] & 0.19 \tnote{(a)}\\
                \tbar~(K) & [66 - 86] & 76 (U), 72 (N) \tnote{(b)}\\
                \hline\hline
            \end{tabular}
        \begin{tablenotes}
            \item[(a)] Solar value
            \item[(b)] Voyager 2
        \end{tablenotes}
        \end{threeparttable}
\end{table}

\section{Inferred atmospheric structures}
\label{sec:results}

\

Here we describe the different atmospheric models obtained from our chemical equilibrium computations, that we ran for both Uranus and Neptune. Since the elemental abundances in Uranus and Neptune are unknown \cite{2022ExA....54..975M}, in this work we kept the same composition settings (metallicity, C/O and S/N ranges) for Uranus and Neptune. The main difference between the two planets is the thermal profile: we constructed thermal profiles over a broad range of \tbar~to consider possible profiles of the atmospheres of both Uranus and Neptune. Due to their similarities, the models presented here can be applied to both planets. 
These  models infer the vertical mass mixing ratio (hereafter MMR) of the main volatiles (with respect to the entire  gas parcel, including all chemical species) and, if present, their associated cloud deck.

It is important to note that determining the physical properties of atmospheric clouds is challenging, as they depend on poorly constrained microphysical processes such as nucleation, particle growth, coagulation, sedimentation, and the availability and nature of condensation nuclei. Recent observations and simulations  \citep{2022JGRE..12707189I,2024PSJ.....5..101G,2024A&A...690A.227C} have begun to place constraints on aerosol properties in the upper troposphere of the ice giants, but a comprehensive description of cloud microphysics and its coupling with atmospheric dynamics is still lacking. \texttt{FastChem} does not explicitly account for these processes, instead it predicts the condensate abundances for a given composition and thermal profile. Therefore, our inferred cloud densities 
can be interpreted as the amount of material expected to reside in the condensed phase under chemical equilibrium conditions, providing an estimate of the condensate reservoir available for cloud formation. As an alternative approach, \cite{2015Icar..245..273W} suggested including dynamical effects by parametrizing a cloud density rate, depending on a prescribed updraft length or duration. For simplicity, we focus on the pressure range at which clouds are predicted to form. The vertical extent of a cloud is not well-defined in chemical equilibrium computations. We therefore define the top of the cloud decks by the pressure at which the cloud density equals 10$^{-4}$ g.l\textsuperscript{-1}. A summary of our results is presented in Table~\ref{tab:modelsummary}.

   \begin{figure}
   \centering
   \includegraphics[width=\hsize]{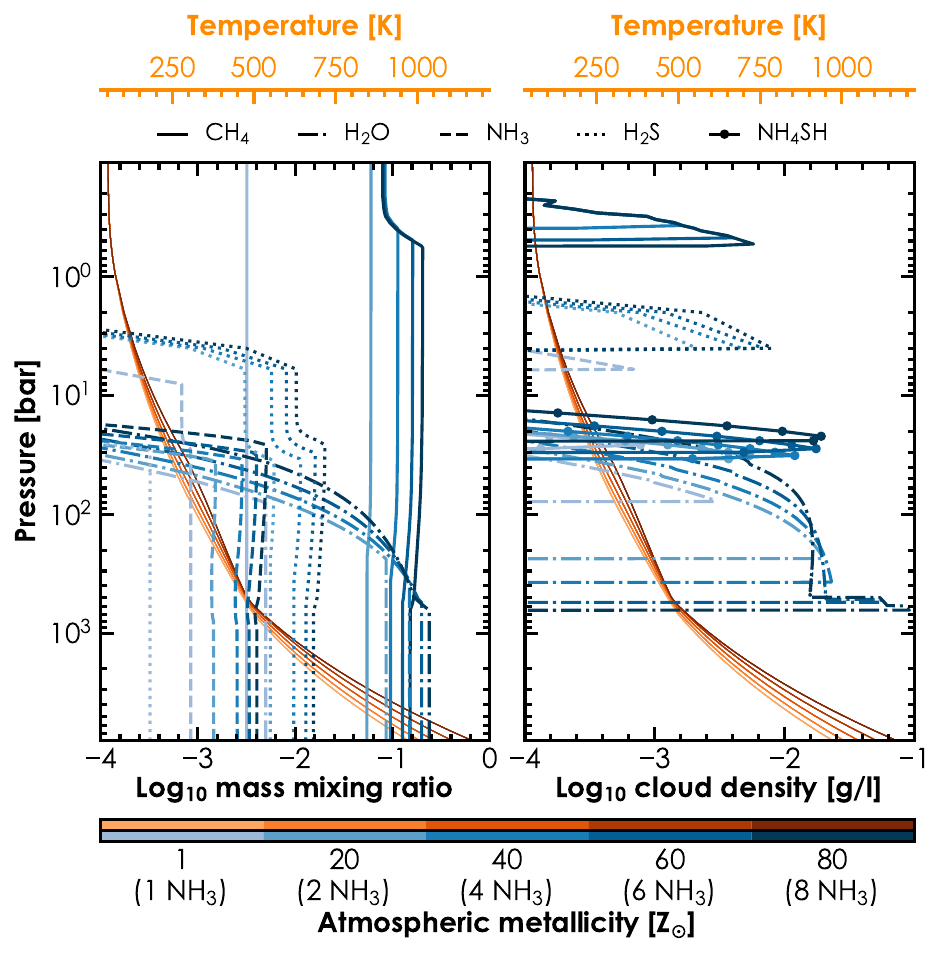}
      \caption{Atmospheric models for different metallicities, with the vertical MMRs (left) and the associated cloud decks (right) for each volatile. Each line style represents a chemical species. The orange lines represent the thermal profiles generated for each metallicity. Here, \tbar~has been set to 76 K.} 
         \label{fig:atm_struct_met}
   \end{figure}

In all of the models the MMRs decrease rapidly as species start to condense, due to the rainout simulation in \texttt{FastChem} that progressively depletes the gas phase. Refractory elements, such as Mg, Si, Ca, or Fe already condense out at very high temperatures and are, therefore, not expected to be present in the upper atmosphere. Since many of these species form oxygen-bearing condensates, such as enstatite (MgSiO$_3$) or forsterite (Mg$_2$SiO$_4$), some of the oxygen in the upper atmosphere is also expected to be depleted. For solar elemental abundances, typically about 20\% of the elemental oxygen is bound in such condensate species \citep{2024MNRAS.527.7263K}. Since NH$_4$SH condensates are formed directly from gaseous H$_2$S and NH$_3$, this species is missing in  the gas phase composition.\\

A solar metallicity atmosphere yields the lowest deep abundances with nearly no condensation, as illustrated in Fig.~\ref{fig:atm_struct_met}. Only a thin H$_2$O cloud appears to form between 30 and 80 bar. However, the cloud thickness is found to increase with metallicity: for the case of Z = 80 Z$_\odot$, water clouds appear in pressures between 20 and 650 bar. At 40 Z$_\odot$, a noticeable CH$_4$ cloud forms above 1 bar and grows to cover a range of 0.2--0.6 bar for 80 Z$_\odot$. This result differs from other atmospheric models such as \citet{2020RSPTA.37890476H}, which predict a methane cloud at the 1 bar level (see Sec.~\ref{sec:discussion} for discussion).  The H$_2$S cloud is found to be the least sensitive to metallicity increase: its position remains similar, between 2 and 5 bar. NH$_4$SH clouds are always found at the top of the water clouds, at pressures of $\sim$ 40 bar. The formation of the NH$_4$SH clouds depletes the gas phase in equal amounts of NH$_3$ and H$_2$S, until the complete desiccation of the least abundant species.  For 1 Z$_\odot$, H$_2$S is the limiting reagent since the solar S/N is 0.19, which explains the inferred absence of an H$_2$S cloud in the model. A NH$_3$ cloud forms between 4 and 6 bars. When the metallicity increases, the NH$_3$ depletion makes H$_2$S the most abundant species in the atmosphere. As a result, the available NH$_3$ is used to form the NH$_4$SH cloud, and  the H$_2$S mixing ratio is reduced by an equivalent amount resulting the condensation of a H$_2$S cloud instead of NH$_3$.

\begin{figure}
   \centering
   \includegraphics[width=\hsize]{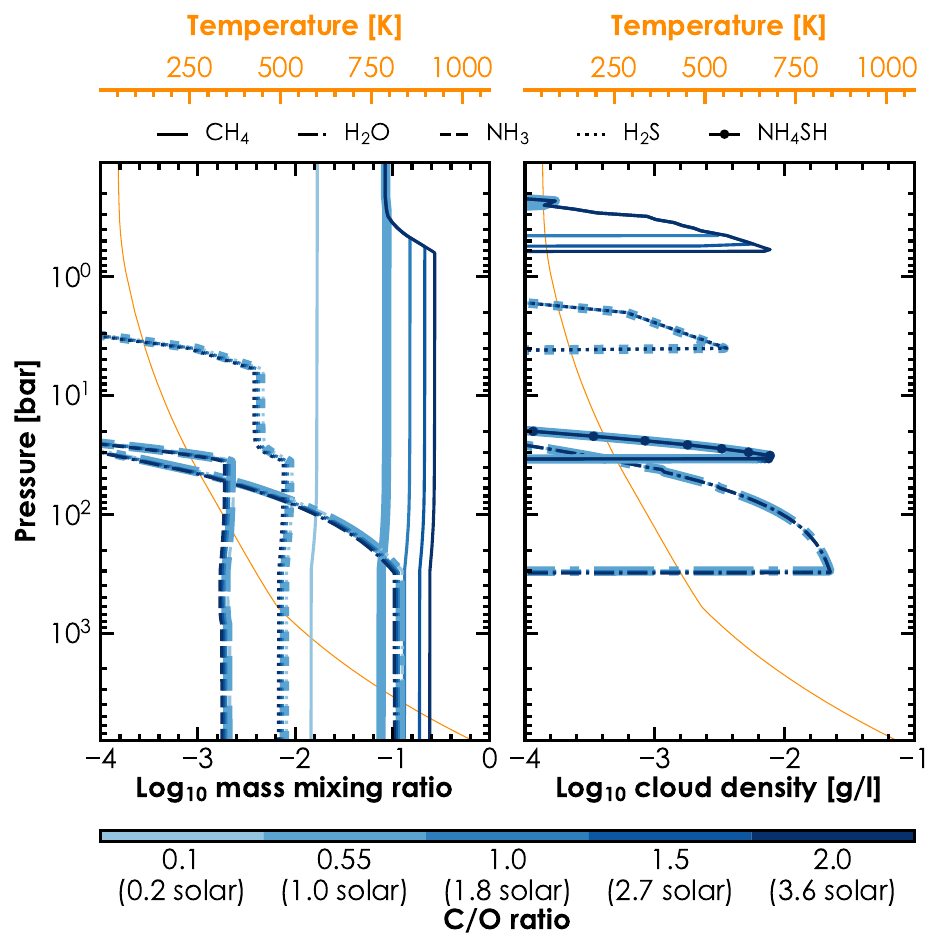}
      \caption{Models for different C/O ratios, for a fixed Z = 30 Z$_\odot$ and \tbar~= 76 K. The thicker lines show the profile for the solar C/O = 0.55.}
         \label{fig:atm_struct_CO}
   \end{figure}

\begin{figure}
   \centering
   \includegraphics[width=\hsize]{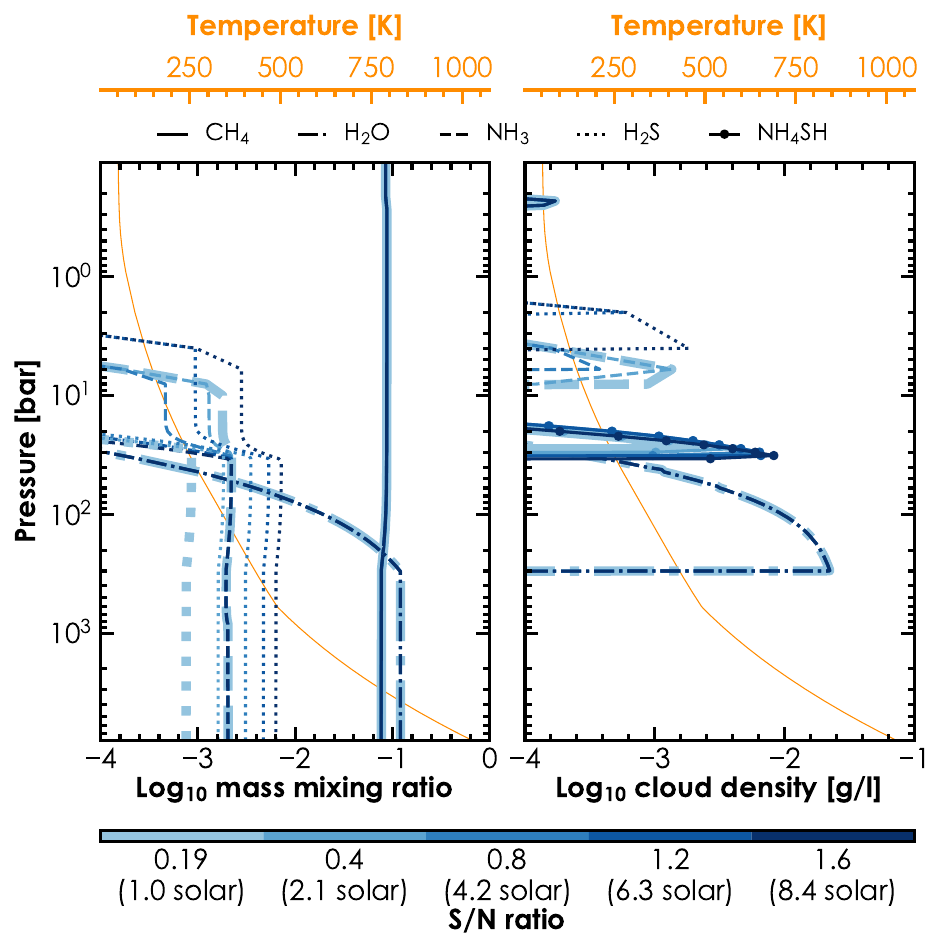}
      \caption{Models for different S/N ratios, for a fixed Z = 30 Z$_\odot$ and \tbar~= 76 K. The thicker lines show the profile for the solar S/N = 0.19.}
         \label{fig:atm_struct_SN}
   \end{figure}

Figures~\ref{fig:atm_struct_CO} \&~\ref{fig:atm_struct_SN} show the atmospheric structures for different assumed C/O and S/N ratios, respectively. For both ratios, we assumed Z = 30 Z$_\odot$ and \tbar~= 76 K. Because we vary the C/O ratio via the carbon abundance, it has a direct impact on the main carbon-bearing species: CH$_4$. Our models show a growth in the methane cloud as the C/O ratio increases. It becomes significantly thick when the ratio reaches unity,  with a pressure range of 0.2--0.45 bar. With a C/O ratio of two, the cloud reaches 0.6 bar.  These pressure levels are lower than those predicted in previous studies, similarly to the high atmospheric metallicity models presented in Fig.~\ref{fig:atm_struct_met}.


Increasing the sulphur abundance strongly affects the  condensation of H$_2$S-NH$_3$. When the S/N ratio is below $\sim$1, NH$_3$ is the most abundant species. Therefore, the condensation of the deep NH$_4$SH cloud completely desiccates the atmosphere in H$_2$S, and NH$_3$ is the only species that condenses (at pressures between 3.6 and 8 bars). With S/N $>$1, H$_2$S condenses instead of NH$_3$ at pressure levels between 1.6--4 bars, which is consistent with observations \citep{2018NatAs...2..420I, 2019Icar..321..550I}. 
The appearance of the NH$_4$SH cloud is insensitive to changes in S/N ratio. Because NH$_4$SH can only exist as condensate, its presence depends on the relative abundances between H$_2$S and NH$_3$. For each S/N ratio, the NH$_4$SH cloud remains at pressures between 17 and 34 bars.


\begin{table*}
\caption{\label{tab:modelsummary} Atmospheric models summary. We report the deep mixing ratios and cloud pressure ranges at the minimum and maximum value of each parameter. For cloud decks, the pressure ranges are defined as [top--base], with the top set to the pressure for which the cloud density equals 10$^{-4}$ g.l\textsuperscript{-1}.}

\centering
\begin{tabular}{lccr}
\hline\hline
Species & Deep mixing ratio & Cloud deck range (bar) & Comments\\ 
 & [Min., Max.] & [Range min.] to [Range max.]\\
\hline
\textit{Metallicity (Z$_\odot$)} &  &  & \textit{For [1, 80] Z$_\odot$ and \tbar~= 76 K}\\
CH$_4$ & [3.16 $\times10^{-3}$, 1.51 $\times10^{-1}$] & \textit{n/a} to [0.20--0.55] & No significant cloud below 40 Z$_\odot$\\
NH$_3$ & [8.44 $\times10^{-4}$, 4.03 $\times10^{-3}$] &  [4.00--6.00] to \textit{n/a} & No cloud for Z $>$ 1 Z$_{\odot}$\\
H$_2$S & [3.23 $\times10^{-4}$, 1.54 $\times10^{-2}$] & \textit{n/a} to [1.60--4.00] & No cloud for Z = 1 Z$_{\odot}$ \\
H$_2$O & [5.00 $\times10^{-3}$, 2.38 $\times10^{-1}$] & [30--80] to [15--650]\\
NH$_4$SH & \textit{n/a} & [20--30] to [15--25]\\

\hline
\textit{C/O ratio} & & & \textit{For [0.1, 2.0], Z = 30 Z$_\odot$ and \tbar~= 76 K}\\
CH$_4$ & [6.74 $\times10^{-2}$, 2.37 $\times10^{-1}$] & \textit{n/a} to [0.22--0.60] & No significant cloud below C/O = 1 \\
NH$_3$ & 1.81 $\times10^{-3}$ & \textit{n/a} \\
H$_2$S & 7.06 $\times10^{-3}$ & 1.60--4.00 \\
H$_2$O & 1.06 $\times10^{-1}$ & 25--300 \\
NH$_4$SH & \textit{n/a} & [20--35]\\

\hline
\textit{S/N ratio} & & & \textit{For [0.19, 1.6], Z = 30 Z$_\odot$ and \tbar~= 76 K}\\
CH$_4$ & 7.64 $\times10^{-2}$ & \textit{n/a} & No significant cloud for Z = 30 Z$_\odot$\\
NH$_3$ & 2.00 $\times10^{-3}$ & [3.60--8.00] to \textit{n/a} & No cloud for S/N $>$ 0.8\\
H$_2$S & [7.57 $\times10^{-4}$, 6.33 $\times10^{-3}$] & \textit{n/a} to [1.68--4.00] & No cloud for S/N $<$ 0.8\\
H$_2$O & 1.20 $\times10^{-1}$ & 25--300\\
NH$_4$SH & \textit{n/a} &[20--35] to [25--42]\\

\hline
\textit{\tbar~(K)} & & & \textit{For [66, 86] K and Z = 30 Z$_\odot$}\\
CH$_4$ & 7.62 $\times10^{-2}$ & [0.32--0.75] to \textit{n/a} & No significant cloud for T $>$ 70 K\\

NH$_3$ & 2.03 $\times10^{-3}$ & \textit{n/a} \\
H$_2$S & [6.10 $\times10^{-3}$, 7.77 $\times10^{-3}$]& [2.60--8.00] to [1.20--2.10]\\
H$_2$O & 1.20 $\times10^{-1}$ & [40--510] to [17--180]\\
NH$_4$SH & \textit{n/a} & [30--52] to [12--23]\\

\hline\hline
\end{tabular}
\end{table*}

\begin{figure}
   \centering
   \includegraphics[width=\hsize]{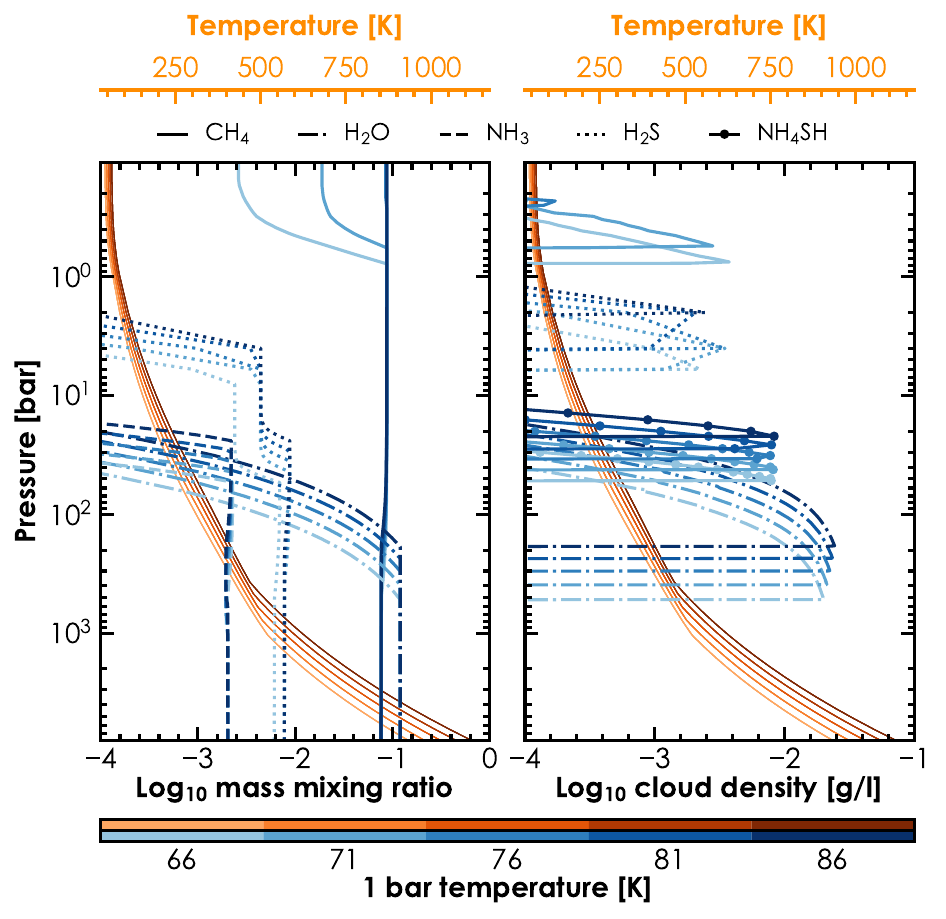}
      \caption{Models for different \tbar, for a fixed Z = 30 Z$_\odot$.}
         \label{fig:atm_struct_temp}
   \end{figure}

Due to the expected (moist) adiabatic nature of the atmosphere, the assumed 1-bar temperature has an impact on the whole profile as well as cloud condensation, as demonstrated in Fig.~\ref{fig:atm_struct_temp}. A difference of $\sim$5 K in \tbar~leads to a difference of $\sim$25 K at 100 bars and $\sim$40 K or more below 1000 bars, for a fixed metallicity of 30 Z$_\odot$. These differences are sufficient to significantly affect the condensation level of clouds. A decrease in \tbar~leads to the appearance of thicker cloud decks at lower pressures. For \tbar~= 66 K, the water clouds appear at pressures of 40--500 bar, compared to 20--200 bar for \tbar~= 86 K. As a reference, in our baseline model with \tbar~= 76 K, the water clouds appear at pressures of 35--300 bar. Since H$_2$S and, even more importantly, CH$_4$, condense at lower temperatures, their respective cloud decks are extremely sensitive to temperature changes. When \tbar~= 66 K, the H$_2$S clouds appear at 2.5--8 bar in comparison to 1--2 bar when \tbar~= 86 K.  Temperature changes have an even more drastic impact on CH$_4$, as no significant cloud deck has been produced beyond \tbar~= 76 K. Only models with \tbar~of 66 K and 70 K predicted CH$_4$ condensation, with clouds appearing at pressure ranges of 0.3--0.8 and 0.25--0.6 bar, respectively. Again, these results correspond to a case with an atmospheric metallicity of Z = 30 Z$_\odot$. More enriched atmospheres could have thicker CH$_4$ clouds despite higher temperatures.\\

\begin{figure*}
\centering
\includegraphics[width=0.8\hsize]{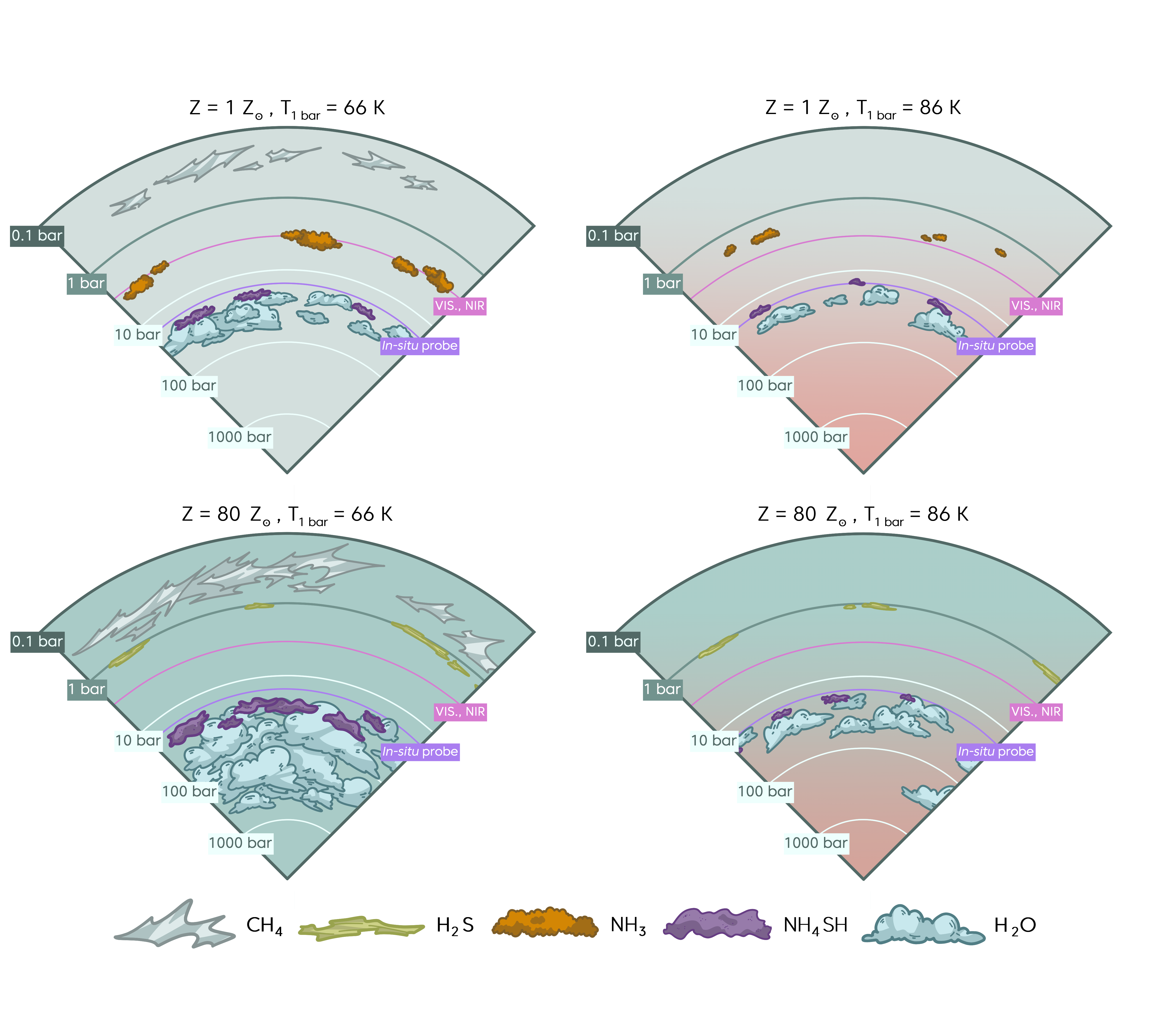}
    \caption{Extreme cases of possible atmospheric structure scenarios derived from our chemical equilibrium computations. Approximate pressure levels reachable by remote observations (in visible and near-infrared wavelengths) and a descending probe are indicated in pink and violet, respectively.}
        \label{fig:atmcases}
\end{figure*}

Overall, our results show the diversity of possible atmospheric structures and compositions. We found that increasing the metallicity and lowering the 1-bar temperature increase the deep mixing ratios and nurture condensation, in a chemical equilibrium framework. We highlight four extreme cases of possible solutions, as illustrated in Fig.~\ref{fig:atmcases}: 

\begin{enumerate}
    \item[$\bullet$]\textbf{Cold, heavy-element-poor atmosphere (1 $Z_{\sun}$, 66 K).} There is a competing tension between a low \tbar~which acts in favour of condensation, and a low Z which reduces it.\\
    \item[$\bullet$]\textbf{Warm, heavy-element-poor atmosphere (1 $Z_{\sun}$, 86 K).} Combined high \tbar~and low Z drastically reduce condensation. The atmosphere exhibits nearly no clouds.\\
    \item[$\bullet$]\textbf{Cold, heavy-element-rich atmosphere (1 $Z_{\sun}$, 66 K).} Both the low \tbar~and high Z act in favour of condensation. Thick clouds are produced in the atmosphere.\\
    \item[$\bullet$]\textbf{Warm, heavy-element-rich atmosphere (1 $Z_{\sun}$, 86 K).} There is a competition between a high \tbar~which acts against condensation, and a high Z which acts in favour.
\end{enumerate}

\section{Discussion}
\label{sec:discussion}
\subsection{Comparisons with previous models}
\label{subs:modelscomp}

Above, we presented the inferred atmospheric structure of Uranus for a range of input parameters and model assumptions. Only a few similar models exist for ice giants, and we compared our findings with those of \citet{2020RSPTA.37890476H} (hereafter H+20) in Table~\ref{tab:comp_hueso}. Since the latter accounts for a 20 Z$_\odot$ metallicity atmosphere, we used our equivalent model, and we can highlight a few differences in the inferred deep mixing ratios and cloud deck bases.

\begin{table}
    \centering
        \caption{\label{tab:comp_hueso} Comparison between our 20 Z$_\odot$ model and the one from \citet{2020RSPTA.37890476H}, accounting for 20 solar H$_2$O. In both cases, MMRs are computed with respect to the whole gas parcel.}
        
        \begin{tabular}{lcc}
            \hline\hline
            Species & This work & H+20\\
            & (20 Z$_\odot$) & \citep{2020RSPTA.37890476H}\\
            \hline
            \textit{Deep mixing ratio} &  & \\
            CH$_4$ &5.40 $\times10^{-2}$ & 6.74 $\times10^{-2}$\\
            NH$_3$ & 1.45 $\times10^{-3}$ & 1.18 $\times10^{-3}$\\
            H$_2$S & 5.55 $\times10^{-3}$ & 7.11 $\times10^{-3}$\\
            H$_2$O & 8.60 $\times10^{-2}$ & 1.34 $\times10^{-1}$\\
            NH$_4$SH & \textit{n/a} & \textit{na}\\

            \hline
            \textit{Cloud deck base (bar)} &  & \\
            CH$_4$ & - & 1.00 \\
            NH$_3$ & \textit{n/a} & \textit{n/a} \\
            H$_2$S & 4.15 & 7.00  \\
            H$_2$O & 240 & 500 \\
            NH$_4$SH & 35 & 50\\

            \hline\hline
        \end{tabular}
\end{table}

Our models generally predict less efficient condensation with thinner clouds, especially for H$_2$O and CH$_4$. If the water cloud is simply thinner in our model, methane does not condense for a 20 Z$_\odot$ metallicity. CH$_4$ cloud forms at higher metallicities, from 40 Z$_\odot$, and the altitude is systematically higher than in previous models (between 0.1 and 1 bar compared to 1--2 bar). In that regard, our computations are closer to older models such as in \citet{2005SSRv..116..121A}. Our inferred atmospheric structure seems to have slightly lower deep mixing ratios in the gas phase. A possible explanation is that the gas phase in \texttt{FastChem} includes about 500 species, whereas the atmospheric composition in the H+20 model is limited to the mentioned volatile species. Because mixing ratios are calculated with respect to the whole gas parcel, a more diverse atmosphere can yield lower individual mixing ratios. However, this argument alone is insufficient for explaining the differences, as only a few species have mixing ratios between 10$^{-5}$ and 10$^{-4}$. Another reason is that, if both models are based on solar abundances from \cite{2009ARA&A..47..481A}, we used present-day abundances, while H+20 uses protosolar ones with slightly higher heavy-element abundances, especially for C and O.

Cloud formation is highly model-dependent. Many chemical equilibrium models trace their origins to the framework developed by \cite{1973Icar...20..465W}, which employs empirically and experimentally derived formulations to calculate thermodynamic properties such as vapour pressures, saturation mixing ratios, and heat capacities. Since then, numerous studies have expanded upon this foundation, advancing the understanding of cloud formation processes in the atmospheres of giant planets \citep[e.g.,][and references therein]{1988JAtS...45.2066C, 2001ApJ...556..872A, 2005SSRv..116..121A, 2015Icar..245..273W, 2024PSJ.....5..101G}.


In this study, the only thermodynamic parametrization required for \texttt{FastChem} is the temperature dependence of the mass action constant for each species. Most of the thermodynamic data used are based on the NIST JANAF tables. As discussed in the appendix of \cite{2024MNRAS.527.7263K}, some condensate data, such as those for H$_2$O and CH$_4$, are taken from other sources. The use of different thermodynamic datasets may explain the discrepancies between the results of this work and those of \cite{2020RSPTA.37890476H} regarding condensate properties and cloud deck bases. Furthermore, as mentioned in Sec.~\ref{sec:results}, the calculation of \texttt{FastChem} represents a condensate reservoir at equilibrium, while estimating actual cloud particle density would require the inclusion of cloud microphysics. Since \cite{2022JGRE..12707189I} put constraint on aerosol properties in the upper troposphere, recent studies have considered more complex processes. For example, \cite{2024PSJ.....5..101G} used heat transport and continuity equations to show that cloud particle density does not depend on the composition of the atmosphere, but on the heat flux and lifetime of the cloud. They inferred the presence of two CH$_4$ cloud decks near 1 and 0.4 bar, as well as a H$_2$S cloud between 5 and 2 bars, with a cloud density varying between 10\textsuperscript{-7} and 10\textsuperscript{-6} kg.m\textsuperscript{-3} (or g.l\textsuperscript{-1}). \cite{2025A&A...694A..81T} investigated the formation, evolution, and precipitation of CH$_4$ cloud droplets, and find cloud decks around 1 bar, with cloud densities reaching nearly 10\textsuperscript{-6} kg.m\textsuperscript{-3}. Overall, our models predict pressure range similar to these recent works, but our cloud densities are much higher. This demonstrates the limitation of chemical equilibrium calculations. However, we note that the more sophisticated models are limited to the upper troposphere while our chemical models include larger atmospheric regions.

\subsection{{Comparisons with observations}}
\label{subs:obscomp}

Previous observations identified the presence of a main cloud deck between 1 and 3 bar \citep{2008DPS....40.5007S, 2012Icar..220..694S, 2019Icar..317..266S}, however, inferring the cloud's composition is challenging. The detection of H$_2$S in Uranus' upper atmosphere  \cite{2018NatAs...2..420I} suggests that this cloud might be composed of hydrogen sulphide \cite{2018NatAs...2..420I}. The  observations were consistent with H$_2$S volume mixing ratio (VMR) between 0.4-0.8 ppm above the cloud deck, and between 1--2 $\times10^{-5}$ below. Comparable values have been found in Neptune's atmosphere \citep{2019Icar..321..550I}. \cite{2022JGRE..12707189I} reanalysed data from various space and ground-based instruments (HST/STIS, IRTF/SpeX, Gemini/NIFS), and inferred the presence of two main clouds down to a few bars. They found a CH$_4$ cloud at pressures of 1--2 bars  and a H$_2$S cloud at pressures of 5--7 bar. Our models predict similar results for H$_2$S while for CH$_4$ we infer condensation at higher altitudes (Figs.~\ref{fig:atm_struct_met} to \ref{fig:atm_struct_temp}).
When converting our MMRs into VMRs, we find H$_2$S abundances between 2.2 $\times10^{-5}$ for 1 Z$_{\sun}$ and 1.18 $\times 10^{-3}$ for 80 Z$_{\sun}$, below the cloud base at $\sim$3--4 bar. Above, the VMRs decreases rapidly ($10^{-6}$--$10^{-8}$). Methane can also be observed in the upper troposphere. \cite{2019Icar..317..266S} monitored latitudinal variations of CH$_4$, and investigated the nature of the main cloud deck between 1.1 and 3.3 bar. Deep VMRs between 2.7 and 3.5 $\times 10^{-2}$ were reported. Our comparable models lie between 40 and 80 Z$_{\sun}$, with respective VMRs between 2 and 3.7 $\times 10^{-2}$. The inferred aerosol properties could correspond to a hydrogen sulphide cloud between 1.1 and 3.3 bar. In this case, the H$_2$S VMR would be greater than 30 ppm at the base of the cloud and below. Our VMRs and cloud altitudes are also consistent with those inferred by observations.

Similar results for chemical species abundances can be obtained by individually changing the elemental ratios, instead of globally increasing the atmospheric metallicity. Radio and microwave interferometric observations of Uranus and Neptune can sample deep structures at tens of bars \citep{2019AJ....157..251T, 2021PSJ.....2..105T, 2021PSJ.....2....3M}, but their limited resolution makes it difficult to produce specific measurements of abundances. In these wavelengths, different species can contribute to the opacity leading to degeneracies in retrievals. Nevertheless, these observations reveal deep latitudinal differences that clearly illustrate the complexities of a single model to characterize the entire planetary atmosphere. 

\subsection{Model limitations}
\label{subs:limit}

The results presented in this paper are based on the assumption that the atmospheres are in thermochemical equilibrium. In reality, however, a variety of processes can  drive the atmosphere away from equilibrium, including vertical mixing, photochemistry, cloud microphysics, and atmospheric dynamics \citep{2020RSPTA.37890477M, 2026arXiv260604510C}. Furthermore, most equilibrium calculations are inherently one-dimensional and do not consider the spatial and temporal variability expected in these planets. Such effects could be important in the cold atmospheres of the ice giants, where chemical and transport timescales can differ substantially and where several key atmospheric properties remain poorly constrained.

In addition, non-equilibrium condensation processes and deviations from saturation can modify the atmospheric structure \citep{2022ExA....54.1027G, 2021PSJ.....2....3M, 2024PSJ.....5..101G, 2024A&A...690A.227C}. Such processes include cloud microphysics, background gas supersaturation or subsaturation.
We also note that our model assumes no meridional and vertical wind shear and therefore does not account for these dynamical sources of mixing.  In reality, these processes would affect cloud condensation and their properties. Investigating the potential impact of these effects on the predicted cloud structure would require multidimensional dynamical simulations \citep{2024PSJ.....5..101G, 2024A&A...690A.227C, 2024A&A...686A.131L}.

Despite these limitations, establishing the equilibrium state of the atmospheres of Uranus and Neptune remains essential for their characterization  step toward understanding giant planet atmospheres \citep[e.g.,][and references therein]{2022ExA....54.1027G}. 
Nevertheless, a more comprehensive modeling approach that includes the relevant physical processes and boundary conditions affecting the inferred atmospheric structure of these planets is highly desirable. We hope to explore such extensions in future studies.

\subsection{Convection inhibition and radiative layers}
\label{subs:convinhibit}

\begin{figure}
\centering
\includegraphics[width=\hsize]{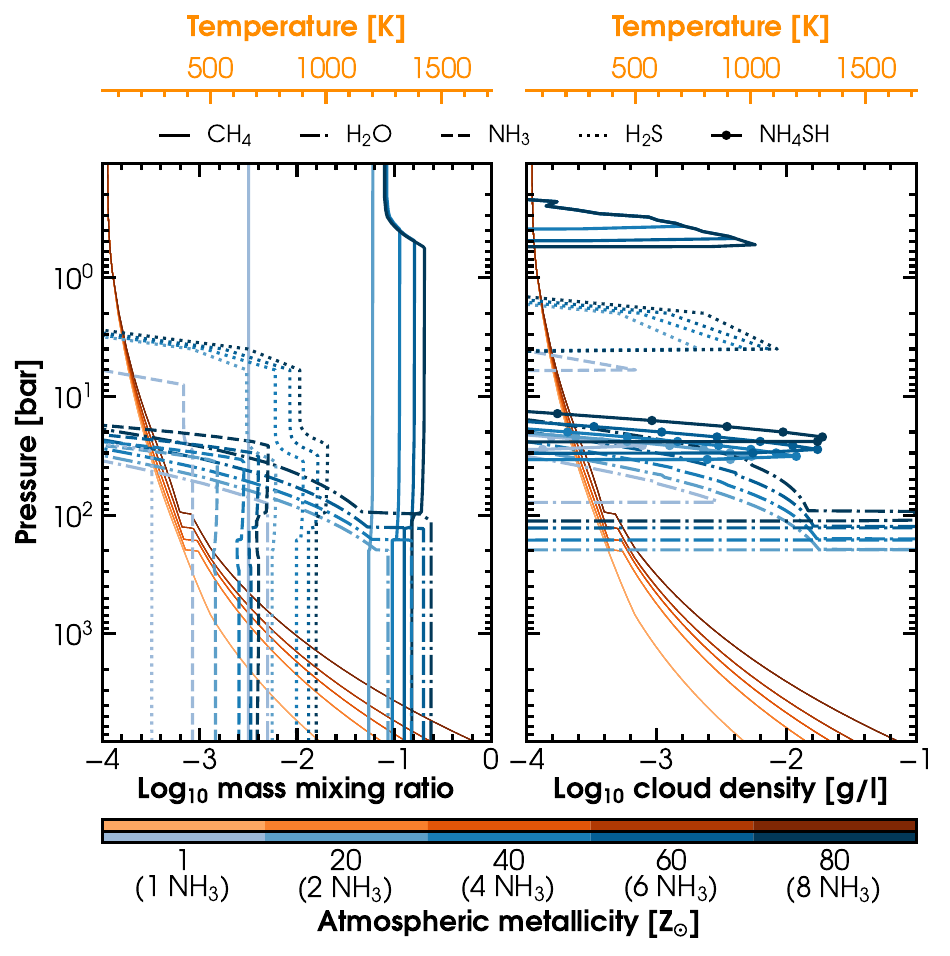}
    \caption{The same profiles as Fig.~\ref{fig:atm_struct_met}, with an inhibition of convection and the formation of a radiative layer at the base of the water cloud deck.}
        \label{fig:PTprofs_conv_inhib}
\end{figure}

As discussed in Sec.~\ref{subs:pt_profile}, composition and thermal profile are interconnected. The molecular weight in a cold, hydrogen-rich atmosphere increases proportionally with the metallicity, and can exhibit sharp gradients between the upper and deeper layers. Such gradients stabilize the atmosphere against convection, and when the vapor abundance reaches a critical value, convection is completely inhibited \citep{1995Sci...269.1697G, 2017A&A...598A..98L, 2022ExA....54.1027G}. An inhibition criterion $\xi_{inh}$ can be added in the computation of thermal profiles, as follows:
\begin{equation}
    \xi_{inh} = \dfrac{\omega M_v L}{RT}q_s, 
\end{equation}
where $\omega = (M_v - M_d)/M_v$ is the reduced mean molar mass difference between the vapor and dry gas phases. Convection is inhibited when $\xi_{\mathrm{inh}} > 1$, leading to the formation of a radiative (stable) layer at the base of the cloud deck. The temperature then follows the radiative gradient $\nabla_r$ (Eq.~\ref{eq: rad-grad}), producing a temperature jump at the cloud base. Fig.~\ref{fig:PTprofs_conv_inhib} illustrates this effect for water. Because water reaches its critical point at much lower pressures and condensation ceases beyond it, the extension of the water cloud is reduced compared to standard adiabatic cases. Given the expected heavy-element enrichment in Uranus and Neptune, convection may also be inhibited by other species such as CH$_4$. We note that this convection-inhibition scenario is the only case where the internal heat flux is considered (see Eq.~\ref{eq: rad-grad}). Otherwise, the atmospheric models are  disconnected from the planetary interior.

\subsection{Implications for future measurements}
\label{subs:futuremeasurements}
Future observations are required for a better understanding of the atmospheres and, consequently, the deeper interior.  
Although Uranus and Neptune have already been observed with various instruments until recently \citep[e.g.][]{2023JGRE..12807980I, 2023A&A...674L...3C, 2025GeoRL..5218301M, 2026arXiv260604510C}, due to the technical limitations, the spatial or spectral resolution of such observation is somewhat limited. Interestingly, current and future high-resolution spectrographs, traditionally built and used for exoplanet science such as ESPRESSO or HARPS/NIRPS and future ones such as ANDES \citep{2024SPIE13096E..13M} or RISTRETTO \citep{2022SPIE12184E..1QL} could significantly improve our understanding of the outer Solar System planets. In addition,  JWST is already producing promising results regarding the properties of the atmospheres of Uranus and Neptune above a few bars \citep{2025epsc.conf.1261R, 2025epsc.conf.1210R}. 

For the lower atmosphere, microwave and centimeter observations provide partial information about the lower atmosphere down to a few tens of bars  \citep{2021PSJ.....2....3M, 2019AJ....157..251T, 2021PSJ.....2..105T}. However, the analysis of such observations is strongly dependent on the assumptions regarding the vertical profiles of temperature and composition, and, thus, remote sensing observations do not resolve the degeneracies between composition, temperature and cloud formation. A major step forward in our understanding of the ice giants would require a dedicated orbiter and probe mission. Long-term, spatially resolved observations, together with in-situ measurements, would provide complementary measurements (e.g., gravity field) that can break some of these degeneracies and further constrain the atmospheric structure of the planets \citep{2020SSRv..216...91H}. Key science goals have already been identified and several potential space missions profiles to Uranus and Neptune have been suggested \citep[][for example]{2024SSRv..220...64V, 2025epsc.conf.1453V, 2020SSRv..216...18A, 2022ExA....54..975M, 2024SSRv..220...50M}. Ground-based observations can also play an important role in supporting mission preparation and to complement measurements done from space.


\section{Summary and Conclusions}
\label{sec:conclusions}

In this study, we investigated the atmosphere of Solar System ice giants using \texttt{FastChem} to recover the vertical distribution and cloud decks of CH$_4$, NH$_3$, H$_2$S and NH$_4$SH. We computed such profiles for a broad range of atmospheric metallicities, C/O and S/N ratios, and thermal profiles. This allowed us to explore the sensitivity of chemical equilibrium models to the assumed elemental abundances and atmospheric entropy. Our key findings can be summarized as follows:

\begin{itemize}
    \item[$\bullet$] \textbf{Assumed Metallicity.} An increase in atmospheric metallicity proportionally increases the mixing ratios and thickness of cloud decks. Within our 1 Z$_\odot$ $\leq$ Z  $\leq$ 80 Z$_\odot$ range, we show that the abundances of H$_2$O, CH$_4$ and H$_2$S can increase by a factor of 50. Due to the NH$_3$ "depletion", its abundance increases only by a factor of five. The pressure ranges for the appearance of H$_2$O and CH$_4$ clouds  increase by a factor of 10 and 2.25, with their base reaching 650 and 0.55 bar, respectively. The condensation of NH$_3$ could yield the formation of a cloud between 4 and 6 bars but its existence depends on the S/N ratio and the formation of a deep NH$_4$SH cloud. We note, however, that observations show that the main cloud deck in the atmospheres of both Uranus and Neptune could be made of H$_2$S \citep{2018NatAs...2..420I,2019Icar..321..550I}.\\
    

    \item[$\bullet$] \textbf{Assumed elemental ratios.} Varying the elemental ratios by targeting a given element affects the inferred mixing ratio and cloud thickness for the main species associated with this element. For our  0.1 $\leq$ C/O $\leq$ 2.0 range, the CH$_4$ deep mixing ratio has increased by a factor of  3.5 and the cloud thickness has increased by a factor of 1.65. For 0.19 $\leq$ S/N $\leq$ 1.6, the H$_2$S deep mixing ratio increased by a factor of $\sim$8. The S/N ratio also controls the condensation of H$_2$S and NH$_3$ through the NH$_4$SH cloud, as NH$_3$ can only condense with S/N $\leq$ 1. For S/N larger than unity, H$_2$S becomes more abundant and is the only species that condenses (at pressures between 1.6 and 4 bars).\\ 

    \item[$\bullet$] \textbf{Assumed atmospheric temperature.} Cold, heavy-element-rich atmospheres are expected to bear prevalent clouds, especially for species condensing at very low temperatures. Within our 66 K $\leq$ \tbar~$\leq$ 86 K range, the base of cloud decks have been lifted up from 510 to 180 bar for H$_2$O, 0.75 to 0.6 bar for CH$_4$ and 8 to 2.1 bar for H$_2$S. Cloud decks thickness have also been reduced by a factor of 3, 1.3 and 6 respectively. Species condensing at low temperatures have been the most affected, with CH$_4$ not significantly condensing beyond \tbar~= 71 K for a 30 Z$_\odot$ atmosphere.\\
    
    \item[$\bullet$] \textbf{Non-uniqueness.} Due to the complex nature of Uranus' and Neptune's atmospheres, the lack of data and the degeneracy between atmospheric parameters, a broad range of atmospheric structures are possible, from cold, heavy-element rich and cloudy atmospheres to warmer, heavy-element poor and cloud-free ones.\\
\end{itemize}

We conclude that the atmospheres of Uranus and Neptune remain mysterious, and currently, a wide range of atmospheric compositions and structures is possible. For a long time, chemical equilibrium models have been used as guides to explore the composition and vertical structure of Uranus and Neptune. Such models are limited as they are one-dimensional, and do not include all the physical and chemical processes occurring in the atmospheres that would affect the vertical structure. Nevertheless, chemical equilibrium models can serve as a baseline, and feed more complex (and computationally expensive) models, such as 3D cloud-resolving models \citep[e.g.][using 1D molecular weight and thermal profiles]{2024A&A...690A.227C} or disequilibrium chemistry models \citep[e.g.][]{2018ApJ...862...31T}, to infer a more complete overview of atmospheric composition and dynamics. New observations over a wide range of wavelengths could further constrain the atmospheric properties of Uranus and Neptune. At the same time, it is clear that dedicated space missions to Uranus and Neptune, combined with ground-based observations can help to break local degeneracies between composition, thermal profile, and cloud formation and to connect the atmospheres with the deep interiors.

\section*{Acknowledgments}

This work has been carried out within the framework of the National Centre of Competence in Research PlanetS supported by the Swiss National Science Foundation under grants \texttt{51NF40\_182901}, \texttt{51NF40\_205606}, and \texttt{215634}. R. Hueso was supported by grant \texttt{PID2023-149055NB-C31} funded by MICIU/AEI/10.13039/501100011033 and FEDER, EU.

\section*{Data availability}

The data underlying this article will be shared on reasonable request to the corresponding author. The \texttt{FastChem} code is available on GitHub at \url{ https://github.com/exoclime/FastChem}.



\bibliographystyle{mnras}
\bibliography{fastchem_biblio} 




\appendix




\section{Thermal profiles parametrization}
\label{apdx:ptprofiles}

In this appendix we show the parameters used for the thermal profiles computation. If the method itself and the equations are the same as in \cite{2017A&A...598A..98L}, we include in Table~\ref{tab:ptprofiles_settings} the source and values for all the variables.

\onecolumn

\begin{table*}
    \centering
        \caption{\label{tab:ptprofiles_settings} Thermal profiles settings summary.}
        \begin{threeparttable}
            \begin{tabular}{lll}
            \hline\hline
            Parameter & Unit & Source/Value\\
            \hline
            \textit{Temperature-dependant variables} & &\\
            $p_{s, H_2O}$ & [Pa] & Tetens formula\\
            $L_{H_2O}$ &[J.kg\textsuperscript{-1}] & Estimated with Clausius-Clapeyron equation\\
            $c_{p, H_2O}$ & [J.kg\textsuperscript{-1}.K\textsuperscript{-1}] & NIST Chemistry Webbook\\
            $c_{p, H_2}$ & [J.kg\textsuperscript{-1}.K\textsuperscript{-1}] & See \textbf{\tnote{(1)}}\\
            $c_{p}$ & [J.kg\textsuperscript{-1}.K\textsuperscript{-1}] & See Equation \ref{eq:cp-mix} \\
            $\kappa$ & [kg.m\textsuperscript{-2}] & See \textbf{\tnote{(2)}}\\
            & & \\
            \textit{Constants} & &\\
            $c_{p, He}$ & [J.kg\textsuperscript{-1}.K\textsuperscript{-1}] & $5R/2$\\
            $q_{int}$ & [kg.kg\textsuperscript{-1}] & set to 0.25\\
            $F_{int}$ & [W.m\textsuperscript{-2}] & 0.042 (U), 0.42 (N) \textbf{\tnote{(3)}}\\
            \hline\hline
            \end{tabular}
            
            \begin{tablenotes}
                \item[\textbf{(1)}] \citet{2014A&A...563A..85V}, for fixed 3:1 ortho-to-para 
                \item[\textbf{(2)}] \citet{2013ApJ...775...10V}
                \item[\textbf{(3)}] \citet{2005AREPS..33..493G}
            \end{tablenotes}
            
        \end{threeparttable}

\end{table*}

\begin{figure*}
\centering
\includegraphics[width=0.8\hsize]{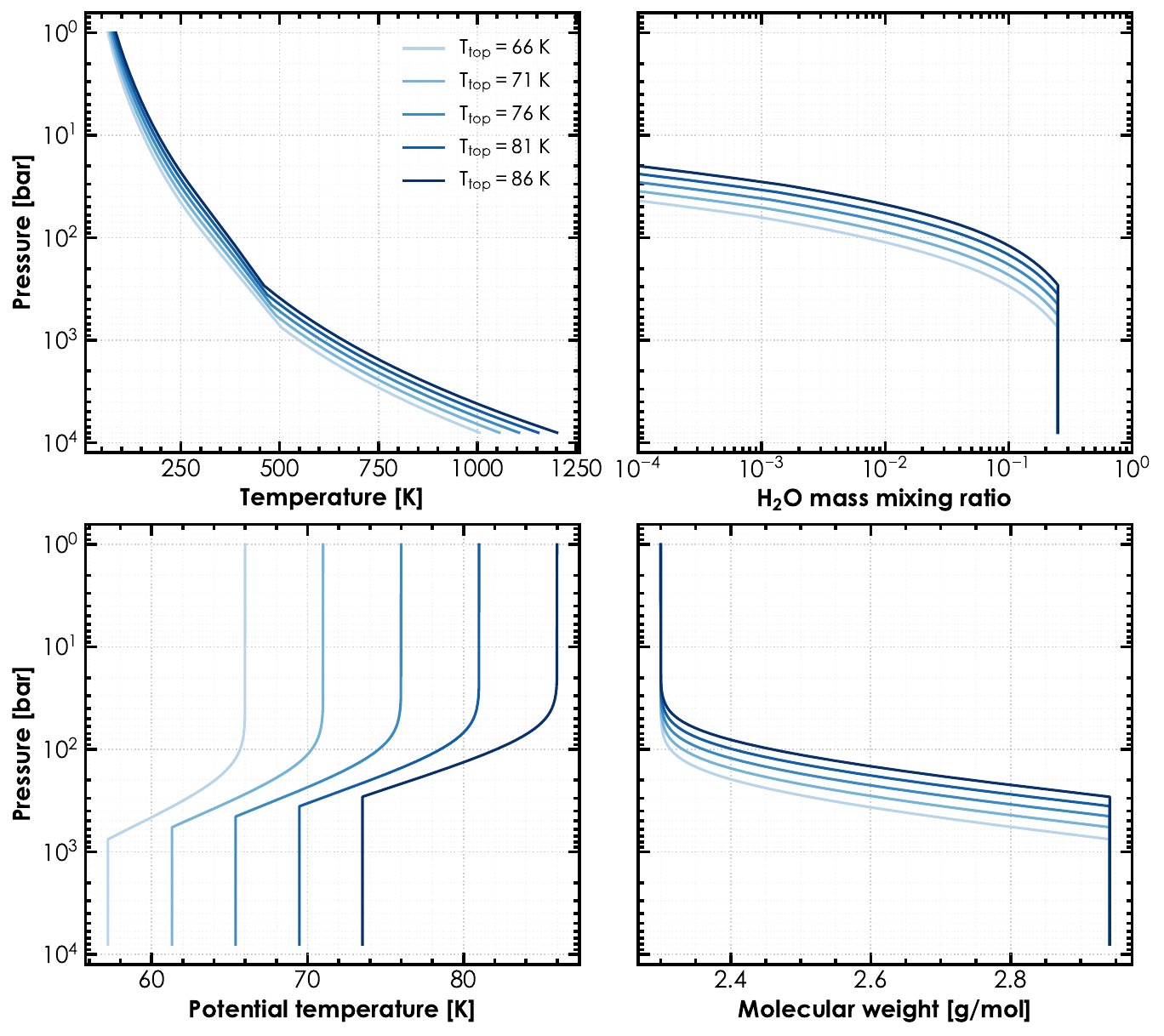}
    \caption{Results of our moist adiabats computations, for a prescribed $q_{int}$ = 0.25. The potential temperature accounts for both the heat capacity and molecular weight of the gas parcel.}
        \label{fig:ptprofs-diag}
\end{figure*}

One can use the saturation pressure of a vapour $p_v$ to recover the saturation mass mixing ratio $q_v$ such as:
\begin{equation}
    q_v = \dfrac{p_vM_{v}}{p_{d}M_{d} + p_vM_{v}} = \dfrac{\rho_{v}}{\rho_d + \rho_v}, 
\end{equation}

where $v$ is the subscript for the vapour (here, H$_2$O), $d$ is the subscript for the dry gas and $M$ indicates a molar mass. When the atmosphere is saturated, we use the saturation pressure $p_{s}$ to recover $q_{s}$. When $q_{s}$ reaches an assumed deep value $q_{int}$, the thermal gradient is expected to transition from a moist to a dry adiabat. In \cite{2017A&A...598A..98L}, $q_{int}$ was a free parameter. In this study we fixed it to 0.25, which corresponds to a median value used by \cite{2017A&A...598A..98L} and compatible with the value inferred by \cite{2020RSPTA.37890476H}. We also show potential temperature (the temperature the gas parcel would have if moved adiabatically to a reference pressure level), saturation mass mixing ratio of water, and molecular weight profiles in Fig.~\ref{fig:ptprofs-diag}. The potential temperature $\theta$ is computed as follows:

\begin{equation}
\theta = T \exp\!\left(-\int_{P_0}^{P} \frac{R}{\mu c_p}\, d\ln P\right) \approx T \left(\frac{P_0}{P}\right)^{\frac{R}{\mu c_p}}
\end{equation}

Where $\mu$ is the mean molecular weight, $c_p$ is the mean specific heat capacity, $R$ the ideal gas constant $P$ the pressure and $P_0$ the reference pressure level. In this work, we took $P_0$ = 1 bar.\\

We calculated the mean specific heat capacity of the gas parcel such as:
\begin{equation}
    c_p(T) = q_d c_{p,d} + q_vc_{p, v} = c_{p, d} + q_v(c_{p, v} - c_{p, d})
    \label{eq:cp-mix}
\end{equation}
For $c_{p, d}$, we considered a dry phase made of 85\% H$_2$ and 15\% He in volume, equivalent to 74\% H$_2$ and 26\% He in mass \citep{1987JGR....9215003C}.

\bsp	
\label{lastpage}
\end{document}